\documentclass[11pt]{article}
\usepackage{amsmath,amssymb,graphicx,amscd}
\usepackage[matrix,arrow,curve]{xy}

\def\im{\Im\hbox{m}}
\def\be{\begin{eqnarray}}
\def\ee{\end{eqnarray}}
\def\nn{\nonumber}
\def\lm{\limits}

\def\pr{\partial}

\def\Tr{{\rm Tr}\,}

\hoffset= - 0.9in         

\title{{\bf S-duality as a $\beta$-deformed Fourier transform} \vspace{.2cm}}
\author{{\bf D.Galakhov}\thanks{{\small
{\it ITEP, Moscow, Russia} and {\it Rutgers University, Piscataway, NJ, USA}};
galakhov@itep.ru, galakhov@physics.rutgers.edu }, {\bf A.Mironov}\footnote{ {\small {\it Lebedev
Physics Institute} and {\it ITEP, Moscow, Russia}}; mironov@itep.ru;
mironov@lpi.ru}, {\bf A.Morozov}\thanks{{\small {\it ITEP, Moscow,
Russia}}; morozov@itep.ru}
\date{ }}
\date{\today}

\begin{document}
\maketitle
\vspace{-5.0cm}

\begin{center}
\hfill FIAN/TD-21/11\\
\hfill ITEP/TH-56/11\\
\end{center}

\vspace{3.5cm}

\begin{abstract}
An attempt is made to formulate Gaiotto's S-duality relations
in an explicit quantitative form.
Formally the problem is that of evaluation of
the Racah coefficients for the Virasoro algebra,
and we approach it with the help of the
matrix model representation of the AGT-related
conformal blocks and Nekrasov functions.
In the Seiberg-Witten limit, this S-duality reduces
to the Legendre transformation.
In the simplest case, its lifting to the level of Nekrasov functions is just the
Fourier transform,
while corrections are related to the beta-deformation.
We calculate them with the help of the matrix model approach and observe that they vanish for
$\beta=1$.
Explicit evaluation of the same corrections from the
$U_q(sl(2))$ infinite-dimensional representation formulas due to B.Ponsot and J.Teshner
remains an open problem.
\end{abstract}

\section{Introduction}

Conformal blocks (CB) naturally arise in consideration of multi-point correlation functions in CFT \cite{BPZ}.
They provide the holomorphic factorization of correlation functions. Note that anomaly free symmetries of the
correlation functions are lost after the holomorphic factorization.
Instead, under the modular transformation the CB are linearly transformed with the help of matrix
of the Racah-Wiegner coefficients, which relates different ways to rearrange the brackets in
associative tensor product \cite{Racah}. A straightforward computation of these fusion relations from basic principles
of CFT is still available only for degenerate representations. Study of this problem in a general context
might reveal some hidden integrable structures in related theories.

Another possible approach \cite{PT} is based on a similar construction for
alternative "conformal blocks" for ${\cal U}_q(sl_2)$. This problem is technically simpler,
and surprisingly the additional deformation parameter $q$ turns to be related to the central charge
$c$ of the Virasoro algebra.

In fact, the Racah-Wiegner coefficients, being a basic notion of representation theory, are important for quite different
subjects. The recently discovered AGT conjecture \cite{AGT,AGTmore} seems to be especially interesting in this context,
since it relates modular properties of the conformal block with a weak-strong coupling S-duality in ${\cal N}=2$
SUSY gauge (SYM) theories. The problem is that the $S$-duality \cite{Sdua} is rather well-understood for the
low-energy effective action  in the Seiberg-Witten (SW) theory \cite{SW1,SW2,Sdua2,Arg}, while the conformal block is
AGT-related with the Nekrasov functions \cite{Nek}
(which describe the $\Omega$-background deformation of the original
SYM theory \cite{Omega}) with two non-vanishing parameters
$\epsilon_1, \epsilon_2\ne 0$, where the $S$-duality
transformations remain unknown. Therefore, the equivalence between the $S$- and modular dualities, which has to be
an essential part of the AGT relation, still lacks any kind of quantitative description.
The purpose of this paper is to initiate consideration of this non-trivial problem.

We begin with the simple and nice functional interpretation of the $S$-duality in the limit of
$\epsilon_1=\epsilon_2=0$, i.e. with the SW theory. In this case,
the SW prepotential $F(a)$
depending on the scalar vacuum expectation value (v.e.v.) of the SYM theory, and its S-dual $F^*(b)$ are related by a simple Legendre transform
\be\label{LT}
F^*(b)=\sup\lm_{a}(F(a)-ab),
\ee
what can be considered as a saddle point approximation to the Fourier integral transform
\be
\exp\left\{\frac{2\pi i}{\epsilon_1\epsilon_2}F^*(b)\right\}=\int da \exp\left\{-\frac{2\pi i a b}{\epsilon_1\epsilon_2}
\right\} \exp\left\{\frac{2\pi i}{\epsilon_1\epsilon_2}F(a)\right\}, \quad \epsilon_{1,2}\rightarrow 0
\ee
The question is how the Fourier transform is deformed when $\epsilon_{1},\epsilon_2
\ne 0$. Technically, the simplest way to
calculate corrections is to use the matrix model description \cite{AGTmamo,MMCB} of the conformal blocks and the
Nekrasov functions. Here the Seiberg-Witten limit and the Fourier transform are just
properties of the spherical limit (the leading order in the genus expansion), and the higher order corrections
can be restored by
well-prescribed procedures like the topological recursion \cite{AMMEO}.
Surprisingly, the Fourier transform
does not acquire corrections in the case of $\beta=1$, at least, in the lowest orders, though there are non-trivial
corrections in the case of $\beta\neq 1$. It would be nice to develop some technique that would allow one
to reproduce these corrections in other approaches.

Another interesting research direction is impled by the fact of appearance of the Racah coefficients
in the modular transformations and in description of the HOMFLY polynomials \cite{HOMFLY}. This is a road to
the most interesting versions of the $3d$ AGT relations \cite{3dAGT}, but it is beyond
the scope of the present paper.

The paper is organized as follows. We briefly describe in sect.2
what are the Racah-Wiegner coefficients
for the Virasoro algebra and their simulation due to \cite{PT}. In sect.3 we review
the AGT relation in the form we need in our further consideration. Formulas for the Racah-Wiegner
coefficients in particular cases, when they are given by the Fourier transform are discussed in sect.4.
At last, in sect.5 we develop the general technique based on matrix model representation and, as an
illustration, calculate a few first corrections to the Fourier transform.
Sect.6 contains comments on $S$-duality transformations in the limit of $\epsilon_2\to 0$.

\section{Racah-Wiegner coefficients for the Virasoro algebra \cite{MS}}
\subsection{Some definitions}

As is well-known, $2d$ conformal field theories can be interpreted in terms of representation theory of the
Virasoro algebra,
\be
\left[L_n,L_m\right]=(n-m)L_{n+m}+\frac{c}{12}n(n^2-1)\delta_{n+m}
\ee
with a non-trivial co-multiplication rule \cite{MS}
\be\label{coproduct}
{\bf \Delta}(L_n)=I\otimes L_n+\sum_{k=-1}^\infty x^{n-k}\left({n+1}\atop{k+1}\right)L_k\otimes I
\ee
which preserves the central charge of the algebra (the central charge of the naive co-multiplication $\tilde {\bf
\Delta}(L_n)=I\otimes L_n+L_n\otimes I$
is twice as large as that of $L_n$).
We denote the co-multiplication with the bold letter in order to
distinguish it from the dimensions of conformal fields.
The primary fields correspond to the highest weight vectors $L_YV_{\Delta}=0$,
$L_0 V_{\Delta}=\Delta V_{\Delta}$, which generate representations (Verma modules)
$R_{\Delta}=\Big\{L_{-Y}V_{\Delta}\Big\}$. The chiral
part of the associative operator product expansion
\be
L_{-Y_1} V_{\Delta_1}(0)\otimes L_{-Y_2} V_{\Delta_2}(x)=\sum\lm_{\Delta,Y}
C^{\Delta,Y}_{\Delta_1,Y_1|\Delta_2,Y_2}L_{-Y}V_{\Delta}(0)\,
x^{\Delta+|Y|-\Delta_1-|Y_1|-\Delta_2-|Y_2|}
\ee
has to be compatible with (\ref{coproduct}), which severely
restricts (almost fixes) the coefficients
$C^{\Delta,Y}_{\Delta_1,Y_1|\Delta_2,Y_2}$.

However, there is more than just the associative algebra structure.
The four-point conformal block is given by the scalar product of a
triple product of representations with fixed
representation in the intermediate channel
and the representation associated with the infinity point
(further we associate it with the fourth point of the CB).
More exactly, one considers the product of
two intertwining operators, $\Phi_{12}^{3}: R_{1}\otimes R_{2}\longrightarrow R_{3}$
which can be combined in two different ways:
$\Phi_{\Delta_1,\Delta_2}^{\Delta}\Phi_{\Delta,\Delta_3}^{\Delta_4}$ and
$\Phi_{\Delta_1,\Delta}^{\Delta_4}\Phi_{\Delta_2,\Delta_3}^{\Delta}$, where
the intermediate representation $R_\Delta$ labels different representations emerging in the
tensor product of two representations.
As usual in representation theory, one demands the co-multiplication to be associative,
which means that the two different products of intertwining operators give just two different
bases of conformal blocks related by an orthogonal $x$-independent matrix $M_{\Delta\Delta'}$
which is called the Racah-Wiegner matrix, i.e. by a linear map
\be\label{racmat}
\Phi_{\Delta_1,\Delta_2}^{\Delta}\Phi_{\Delta,\Delta_3}^{\Delta_4}=
\sum\lm_{\Delta'}M_{\Delta\Delta'}
\Phi_{\Delta_1,\Delta'}^{\Delta_4}\Phi_{\Delta_2,\Delta_3}^{\Delta'}
\ee
Instead of considering these products of intertwining operators, one can
calculate their value on the highest weight vectors.
Since the dual of $R_{\Delta_4}$ enters the answer, it can be written using the scalar product.
This product is called the conformal block
and, thus, there are two essentially different conformal blocks.
They can be presented pictorially as
\be\label{CB}
\xymatrix{
&B_{\Delta}(x)=\Big<\Big( V_{\Delta_1}(0)\otimes V_{\Delta_2}(x)\Big)_{\Delta}
\otimes V_{\Delta_3}(1),
V_{\Delta_4}(\infty)\Big> \ar@{<=>}[dd]^{B_{\Delta}=
\sum\lm_{\Delta'}M_{\Delta\Delta'}B_{\Delta'}}\\
\Big< V_{\Delta_1}(0)V_{\Delta_2}(x)V_{\Delta_3}(1)V_{\Delta_4}(\infty)\Big> \ar[ur]^{I}
\ar[dr]_{II} &\\
& B_{\Delta'}(1-x)=\Big< V_{\Delta_1}(0)\otimes \Big( V_{\Delta_2}(x)
\otimes V_{\Delta_3}(1)\Big)_{\Delta'},
V_{\Delta_4}(\infty)\Big>
}
\ee
In terms of the conformal blocks relation (\ref{racmat}) can be rewritten as
\be\label{Mod_tr}
\boxed{
B_{\Delta}(x)=\sum\lm_{\Delta'}M_{\Delta\Delta'}B_{\Delta'}(1-x)}
\ee
Our goal is to study this map within the context of AGT. On one side, $M_{\Delta\Delta'}$
describes modular transformations of the conformal block. On another side, it is a deformation of
the Legendre transform (\ref{LT}) to $\epsilon_1,\epsilon_2\ne 0$.

\subsection{Various modular transformations}

It is worth noticing that relation (\ref{Mod_tr}) does not exhaust all possible relations between
different types of conformal blocks. Indeed, one could change the order of vertex operators
inside the brackets and rearrange the brackets in different way in the final scalar product.
In fact, one can construct all possible different relation by two independent transformations
\be
\hat {S}:\ \ \ \ \ x\to 1-x,\ \ \ \ \hat {T}:\ \ \ \ \ x\to {x\over x-1}
\ee
The first transformation connects the following two conformal blocks
\unitlength 1.5mm 
\linethickness{0.4pt}
\be
\begin{picture}(30,14)\put(0,0){\line(1,0){30}}
  \put(10,0){\line(0,1){10}}
  \put(20,0){\line(0,1){10}}
  \put(1,2){\makebox(0,0)[cc]{$0,\Delta_1$}}
  \put(6,9){\makebox(0,0)[cc]{$x,\Delta_2$}}
  \put(15,2){\makebox(0,0)[cc]{$\Delta$}}
  \put(16,9){\makebox(0,0)[cc]{$1,\Delta_3$}}
  \put(30,2){\makebox(0,0)[cc]{$\infty,\Delta_4$}}
  \end{picture}\qquad\mathop{\longrightarrow}^{\hat S} \qquad\begin{picture}(30,14)\put(0,0){\line(1,0){30}}
  \put(10,0){\line(0,1){10}}
  \put(20,0){\line(0,1){10}}
  \put(1,2){\makebox(0,0)[cc]{$1,\Delta_3$}}
  \put(6,9){\makebox(0,0)[cc]{$x,\Delta_2$}}
   \put(15,2){\makebox(0,0)[cc]{$\Delta'$}}
  \put(16,9){\makebox(0,0)[cc]{$0,\Delta_1$}}
  \put(30,2){\makebox(0,0)[cc]{$\infty,\Delta_4$}}
  \end{picture}
\ee
This $S$-duality transformation looks especially simple in terms of the
"the effective coupling constant" ${\cal T}$ (see s.3.1 below for
the explanation of the terminology) in the SW limit:
\be
\hat {S}:\quad {\cal T}\rightarrow -\frac{1}{\cal T}
\ee
The second generator, $\hat T$ which describes the second modular transformation
(the two being enough to give rise to the whole modular group),
in this case looks like
\be
\hat {T}:\quad {\cal T} \rightarrow {\cal T}+1
\ee
This transformation connects the following conformal blocks
\unitlength 1.5mm 
\linethickness{0.4pt}
\be
\begin{picture}(30,14)\put(0,0){\line(1,0){30}}
  \put(10,0){\line(0,1){10}}
  \put(20,0){\line(0,1){10}}
  \put(1,2){\makebox(0,0)[cc]{$0,\Delta_1$}}
  \put(6,9){\makebox(0,0)[cc]{$x,\Delta_2$}}
   \put(15,2){\makebox(0,0)[cc]{$\Delta$}}
  \put(16,9){\makebox(0,0)[cc]{$1,\Delta_3$}}
  \put(30,2){\makebox(0,0)[cc]{$\infty,\Delta_4$}}
  \end{picture}\qquad\mathop{\longrightarrow}^{\hat T} \qquad\begin{picture}(30,14)\put(0,0){\line(1,0){30}}
  \put(10,0){\line(0,1){10}}
  \put(20,0){\line(0,1){10}}
  \put(1,2){\makebox(0,0)[cc]{$0,\Delta_1$}}
  \put(6,9){\makebox(0,0)[cc]{$x,\Delta_2$}}
 \put(15,2){\makebox(0,0)[cc]{$\Delta'$}}
  \put(16,9){\makebox(0,0)[cc]{$\infty,\Delta_4$}}
  \put(30,2){\makebox(0,0)[cc]{$1,\Delta_3$}}
  \end{picture}
\ee

One can also use "the bare coupling constant" $\tau_0$: $x=e^{2\pi i\tau_0}$
(see s.3.1), though in these terms the same transformations look quite ugly:
\be
\hat {S}:\ \ \ \ \ e^{2\pi i\tau_0}\to 1-e^{2\pi i\tau_0},\ \ \ \
\hat {T}:\ \ \ \ \ e^{2\pi i\tau_0}\to {e^{2\pi i\tau_0}\over e^{2\pi i\tau_0}-1}
\ee

In the generic $\Omega$-background the conformal block transforms non-trivially
only w.r.t. the first transformation, $\hat {S}$, while the $\hat {T}$-transformation
just gives rise to a trivial factor:
\be
B_\Delta\Big(\Delta_1,\Delta_2,\Delta_3,\Delta_4|x\Big)=(-1)^\Delta(1-x)^{2\Delta_2}
B_\Delta\Big(\Delta_1,\Delta_2,\Delta_4,\Delta_3|x\Big)
\ee
This is because $\hat {T}$ interchanges the points $1$ and $\infty$ and does not affect
the point $x$.
At the same time, (\ref{Mod_tr}) is absolutely non-trivial transformation, so
we mostly concentrate on it.

We shall also consider modular transformations of the one-point conformal block on a
torus which depends on the modular parameter of the torus $\tau_0$. These modular
transformations are generated by two independent transformations
\be
\hat {S}:\ \ \ \ \ \tau_0\to -{1\over\tau_0},\ \ \ \ \hat {T}:\ \ \ \ \ \tau_0\to \tau_0+1
\ee
In this case also only the $\hat S$-transformation is non-trivial, while
the $\hat {T}$-transformation
just gives rise to a phase factor:
\be
B_\Delta\Big(\Delta_{ext}|\tau_0+1\Big)=\exp\Big\{2\pi i\Big(\Delta-{c\over 24}\Big)\Big\}
B_\Delta\Big(\Delta_{ext}|\tau_0\Big)
\ee
We return to discussion of the whole modular ($S$-duality) group in s.6.

\subsection{Racah-Wiegner coefficients for Virasoro from ${\cal U}_q(sl_2)$-representations}

The problem of constructing the Racah-Wiegner matrix for the Virasoro algebra
was solved in the case of degenerate representations \cite{MS}, though in generic situation it
is quite involved. Instead, in \cite{PT} B.Ponsot and J.Teschner studied the Racah-Wiegner matrix
for specific infinite-dimensional representations of the algebra $U_q(sl_2)$ and suggested that
it is equal to that for the Virasoro case.

In fact, the modular transformation was explicitly described in \cite{PT} in two cases, AGT-related
to $SU(2)$ SYM theory with $N_f=2N_c=4$ matter hypermultiplets or with one adjoint matter multiplet.
The first case is the spherical four-point conformal block \cite{PT}:
\be\label{6}
B_{p}\left(\left.\begin{array}{cc}
                        p_1 & p_2 \\
                        p_3 & p_4
                      \end{array}
\right|x\right)=\int d \mu(p') M_{p p'}\left(\begin{array}{cc}
                        p_1 & p_2 \\
                        p_3 & p_4
                      \end{array}\right)B_{p'}\left(\left.\begin{array}{cc}
                        p_2 & p_3 \\
                        p_4 & p_1
                      \end{array}
\right|1-x\right)
\ee
where
\be
\Delta(p)=\frac{\epsilon^2/4+p^2}{\epsilon_1\epsilon_2},\quad d\mu(p)=4\sinh\left(2\pi\frac{p}{\epsilon_1}
\right)\sinh\left(2\pi\frac{p}{\epsilon_2} \right) dp,\ \ \ \ \ \ \ \epsilon=\epsilon_1+\epsilon_2
\ee
and
\be
M_{p p'}\left(\begin{array}{cc}
                        p_1 & p_2 \\
                        p_3 & p_4
                      \end{array}\right)=\frac{s(u_1)s(w_1)}{s(u_2)s(w_2)}\int\lm_{\mathbb{R}} dt\prod\lm_{i=1}^4\frac{s(t-r_i)}{s(t-q_i)}\\
\begin{array}{cccc}
  r_1=p_2-p_1 & q_1=\epsilon/2-p_4+p_2-p' & u_1=p+p_2-p_1 \\
  r_2=p_2+p_1 & q_2=\epsilon/2-p_4+p_2+p' & u_2=p+p_3+p_4  \\
  r_3=-p_4-p_3 & q_3=\epsilon/2+p &  w_1=p'+p_1+p_4 \\
  r_4=-p_4+p_3 & q_4=\epsilon/2-p & w_2=p'+p_2-p_3\\
\end{array}
\ee
while the second case is the one-point toric conformal block
\be\label{10}
B_p(p_0|{\cal T})=\int d\mu(p')M_{pp'}(p_0)B_{p'}(p_0|-1/{\cal T})\\
\label{toric_MK} M_{pp'}(p_0)=\frac{2^{\frac{3}{2}}}{s(p_0)} \int\lm_{\mathbb{R}} dt \frac{s\left(p'+\frac{1}{2}(p_0+\epsilon)+t\right)s\left(p'+\frac{1}{2}(p_0+\epsilon)-t\right)}{s\left(p'-\frac{1}{2}(p_0+\epsilon)+t\right)s\left(p'-\frac{1}{2}(p_0+\epsilon)-t\right)}e^{4\pi i p t}
\ee
where we used "the quantum dilogarithm" \cite{qdl},
the ratio of two digamma-functions \cite{diga},
\be
\log s(z|\epsilon_1,\epsilon_2)=\frac{1}{i}\int\lm_0^{\infty}\frac{dt}{t}
\left(\frac{\sin 2x t}{2\sinh \epsilon_1 t \sinh \epsilon_2 t}-\frac{x}{\epsilon_1\epsilon_2 t}
\right)=\nn\\
=
\prod_{m,n\geq 0}
\frac{\left(m+\frac{1}{2}\right)\epsilon_1
+\left(n+\frac{1}{2}\right)\epsilon_2 - i z}
{\left(m+\frac{1}{2}\right)\epsilon_1
+\left(n+\frac{1}{2}\right)\epsilon_2+i z}
= \frac{\Gamma_2(\epsilon/2+iz|\epsilon_1,\epsilon_2)}
{\Gamma_2(\epsilon/2-iz|\epsilon_1,\epsilon_2)}
\ee
which possesses a number of periodicity properties
\be\label{per}
s\left(z-\frac{i\epsilon_2}{2}\Big|\epsilon_1,\epsilon_2\right)=2\cosh\left(\frac{\pi z}{\epsilon_1}\right)s\left(z+\frac{i\epsilon_2}{2}\Big|\epsilon_1,\epsilon_2\right)\\
s\left(z-\frac{i\epsilon_1}{2}\Big|\epsilon_1,\epsilon_2\right)=2\cosh\left(\frac{\pi z}{\epsilon_2}\right)s\left(z+\frac{i\epsilon_1}{2}\Big|\epsilon_1,\epsilon_2\right)\\
s\left(z-\frac{i\epsilon}{2}\Big|\epsilon_1,\epsilon_2\right)=4\sinh\left(\frac{\pi z}{\epsilon_1}\right)\sinh\left(\frac{\pi z}{\epsilon_2}\right)s\left(z+\frac{i\epsilon}{2}\Big|\epsilon_1,\epsilon_2\right)
\label{shift_s}\label{s_rel}
\ee
Formulas (\ref{6}), (\ref{10}) and (\ref{toric_MK}) should be considered as
contour integrals around
poles and zeroes of the quantum dilogarithms
and they are quite difficult to use. They can be simplified in some particular cases.

\section{AGT representation for conformal block}
\subsection{Nekrasov partition function}
The complex coupling constant in the Yang-Mills theory is defined in
terms of the standard coupling constant $g$ and the $\theta$-angle
as
\be
{\cal T}=\frac{4\pi i}{g^2}+\frac{\theta}{2\pi}
\ee
The internal
symmetry transformation of the theory ${\cal T}\mapsto{\cal T}+1$ and the
duality map (S-duality) defined by N.Seiberg and E.Witten \cite{SW1},
form the modular group $SL(2,\mathbb{Z})$. It is important to
distinguish between the bare coupling constant $\tau_0$ arising in
the fundamental SYM theory and the effective one, ${\cal T}$ arising in
the low-energy effective action of the general form
\be
S_{eff}=\frac{1}{4\pi}\int d^4 x\,\im\left[\int d^2\theta\,
\frac{\pr\mathcal{F}(A)}{\pr A}\bar{A}+\frac{1}{2}\int d^4 \theta\,
\frac{\pr^2\mathcal{F}(A)}{\pr A^2}W_{\alpha}W^{\alpha}\right]
\ee
as ${\cal T}(a)=\pr^2\mathcal{F}(a)/\pr a^2$, where the modulus $a$ is
essentially the scalar field v.e.v. We are interested in action of
S-duality on both ${\cal T}$ and $\tau_0$. The SW prepotential can be
simply related \cite{Nek} to the LMNS integral \cite{LMNS}
$$
\mathcal{F}=\lim\lm_{\epsilon_{1,2}\rightarrow 0}\frac{\epsilon_1\epsilon_2}{2\pi i }\log \mathcal{Z}_{\rm LMNS}
$$

The LMNS integral is defined for the ${\cal N}=2$ SYM theory on the so called $\Omega$-background parameterized by two
parameters $\epsilon_1$ and $\epsilon_2$. The simplest is the theory with the gauge group $SU(2)$ and
four fundamental matter hypermultiplets with masses $\mu_i$ (the $\beta$-function in this theory is vanishing).
In this case, the LMNS integral
is represented \cite{Nek} by a power series in exponential of the bare coupling constant $\tau_0$
parameterized by pairs of the Young diagrams $Y_1$, $Y_2$
\be
Z_{\rm Nek}=\sum\lm_{Y_1,Y_2} N_{\epsilon_{1,2}}(Y_1,Y_2,\mu_1,\mu_2,\mu_3,\mu_4,a) e^{2\pi i\tau_0 (|Y_1|+|Y_2|)}
\ee
The coefficients $N_{\epsilon_{1,2}}(Y_1,Y_2,\mu_1,\mu_2,\mu_3,\mu_4,a)$
(the  Nekrasov functions)
are rational functions in $\epsilon$'s, the scalar v.e.v. $a$, and the matter hypermultiplet masses $\mu_i$'s.

This series  is known to coincide \cite{AGT,AGTmore,AGTrev} with the four-point conformal block up to a so called
$U(1)$-factor under the following identification of the CFT and SYM data:
\be
\Delta(\alpha)=\frac{\alpha(\epsilon_1+\epsilon_2-\alpha)}{\epsilon_1\epsilon_2}, c=1+6\frac{(\epsilon_1+\epsilon_2)^2}{\epsilon_1\epsilon_2}\\
\mu_1=-\epsilon/2+\alpha_1+\alpha_2,\quad \mu_2=\epsilon/2+\alpha_2-\alpha_1, \quad \mu_3=-\epsilon/2+\alpha_3+\alpha_4, \quad \mu_4=\epsilon/2+\alpha_3-\alpha_4\\
a=\alpha-\epsilon/2,\quad x=e^{2\pi i\tau_0}
\ee
where $\alpha_i$ corresponds to the four external dimensions and $\alpha$ to the intermediate dimension in (\ref{CB}).

It is remarkable that these Nekrasov functions do not possess any specific symmetry,
though the Seiberg-Witten theory
does \cite{SW1,SW2}. Indeed, the Seiberg-Witten theory is symmetric under the transformations
\be
\hat{T}:\quad {\cal T}\rightarrow{\cal T}+1
\\
\hat{S}: \quad {\cal T}\rightarrow -\frac{1}{{\cal T}}\nn
\ee
This symmetry reflects the freedom in choosing the $A$- and $B$-cycles on the spectral curve.
As we shall see, these symmetries can be lifted to the level of the $\epsilon$-deformed prepotential or the
Nekrasov functions. The idea of this identification was presented in \cite{Gaiotto}.
It is also supported by the fact that the description in terms of the SW equations with
contour integrals remains true for $\epsilon_1,\epsilon_2\ne 0$ \cite{MMCB}.

\subsection{Matrix models\label{mamo}}
As is shown in \cite{MMCB} the conformal block and, hence, the Nekrasov functions can be defined in terms of
the $\beta$-ensemble (later on, we often call it just matrix model, though it is literally a matrix model only at $c=1$,
i.e. at $\epsilon_1=-\epsilon_2$):
\be\label{mmint}
Z=\prod_{a<b} (q_q-q_b)^{2\alpha_a\alpha_b\over g} \int_{\gamma_i} dz_i \left(\prod\lm_{j>i} z_{ij}^{2\beta}\right)
\prod\lm_a (z_i-q_a)^{\frac{2b\alpha_a}{g}},\quad g=\sqrt{-\epsilon_1\epsilon_2},\ \ \
\beta=b^2=-\frac{\epsilon_1}{\epsilon_2}
\ee
Here $a,b=1,2,3$, $q_1=0$, $q_2=x$, $q_3=1$ and among the integration contours $\gamma_i$ there are
\be\label{N1}
N_1={1\over b}\Big(\alpha-\alpha_1-\alpha_2\Big)
\ee
segments $[0,q]$ and
\be\label{N2}
N_2={1\over b}\Big(b-{1\over b}-\alpha-\alpha_3-\alpha_4\Big)
\ee
segments $[0,1]$. This partition function satisfies the
Seiberg-Witten equations: the prepotential $F=g^2\log Z$ can be
restored from
\be\label{mmSW}
a=\oint\lm_A \Omega_{g,\beta}, \quad
\frac{\pr F(a)}{\pr a} =\oint\lm_B \Omega_{g,\beta}
\ee
where
$\Omega_{g,\beta}$ is the full (all genus) one-point resolvent of the matrix model,
see the next section. This allows one to lift the SW construction
to the level of the Nekrasov functions and extend the S-duality
transformation to the conformal block. As we shall see further, this
transformation provides exactly the modular transformation.

\section{Modular transformation as Fourier transform}
\subsection{The simplest case and strategy}

To see how the modular transformation can be interpreted in terms of Seiberg -Witten theory
note that, due to the AGT correspondence, the conformal block behaves similarly to the Nekrasov functions,
in particular, in the limit of both $\epsilon_1$ and $\epsilon_2$ going to zero:
\be
B\ \ {\stackrel{\epsilon_{1,2}\rightarrow 0}{\sim}}\  \ \exp
\Big\{\frac{2\pi i}{\epsilon_1\epsilon_2} F_{SW}(a)\Big\}
\ee
Here $F_{SW}$ is the Seiberg-Witten prepotential defined for the effective curve
\be
\oint\lm_A \Omega_{SW} =a,\quad \oint\lm_B \Omega_{SW} =\frac{\pr F_{SW}(a)}{\pr a}
\ee
At the same time, one can make another choice of contours and define another prepotential
\be
\oint\lm_A \Omega_{SW} =-\frac{\pr F^*_{SW}(b)}{\pr b},\quad \oint\lm_B \Omega_{SW} =b
\ee
Consider the simplest example of this construction corresponding to the four-point conformal block with
external fields of zero conformal dimensions, or, in terms of ${\cal N}=2$ SYM theory, with four massless
matter hypermultiplets (note that in the deformed case these masses become proportional to
$\epsilon=\epsilon_1+\epsilon_2$
if the conformal dimensions are still zero). The corresponding SW differential reads
\be
\Omega_{SW}=\frac{u dz}{\sqrt{z(z-x)(z-1)}}
\ee
The cycles are chosen as shown in Fig.1.
\begin{figure}[h]
\begin{center}
\includegraphics[scale=1]{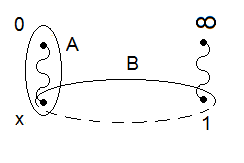}
\end{center}
\label{4_cuts}
\caption{Cycles for the 4-point conformal block}
\end{figure}
One can directly compute both the periods and the corresponding prepotential to obtain
\be
\oint\lm_A \Omega_{SW}=\frac{1}{\pi i}K(x),\, \ \ \ \
\oint\lm_B \Omega_{SW}=\frac{1}{\pi}K(1-x),\, \ \ \ \
F(a)=\frac{i a^2}{2}\frac{K(1-x)}{K(x)}
\ee
It is immediate to observe that the rational transformation $z\mapsto 1-z$ permutes the contours $(A,B)\mapsto(-B,A)$,
so that one obtains the conjugated prepotential with the variable $x$ replaced with $1-x$. In other words,
\be
\boxed{F^*(a,x)=F(a,1-x)}
\ee
This relation works equally well in the generic case even if the masses or the deformation parameters
$\epsilon_{1,2}$ are non-zero. Since the prepotential is directly related to the conformal block, one naturally
associates this permutation of contours with the modular transformation $x\to 1-x$ of the prepotential, i.e.
\be
\boxed{F\mathop{\longrightarrow}^{M_{\Delta,\Delta'}}F^*}
\ee
This statement allows one to construct the corresponding modular matrix in an explicit way.
Indeed, the S-duality transformation that relates the SW prepotential and its conjugated is known to be
nothing but the Legendre transformation
\be
F^*(a,x)=\sup\lm_b \Big(F(b,x)-ab\Big)=F(a,1-x)
\ee
This relation can be understood as the leading (semiclassical) approximation to
\be
\exp\left\{\frac{2\pi i}{\epsilon_1\epsilon_2}F^*(a)\right\}=\int db\,
\exp\left\{\frac{2\pi i}{\epsilon_1\epsilon_2} ab \right\}
\exp\left\{\frac{2\pi i}{\epsilon_1\epsilon_2}F^*(b)\right\}
\ee
at small deformation parameters $\epsilon_{1,2}$. In other words, the asymptotic behavior of the modular matrix is
\be
M_{a,b}\sim \exp\left\{\frac{2\pi i}{\epsilon_1\epsilon_2} ab \right\}
\ee

\subsection{Exactly solvable cases}\label{Ex_cases}

We are going to consider now the cases when the SW approximation
turns out to be almost exact. It happens when the coefficients in
the Zamolodchikov recursive formula \cite{Zam}\footnote{
It was considered within the AGT context in \cite{MMMz,Pog}.} which are in charge
of the "non-classical" part are equal to zero:
\be
B(\Delta_1,\Delta_2,\Delta_3,\Delta_4;\Delta|x) =
\left(e^{\pi i{\cal T}}\right)^{\Delta - \frac{c-1}{24}}x^{\frac{c-1}{24}-\Delta_1-\Delta_2}
(1-x)^{\frac{c-1}{24}-\Delta_2-\Delta_3}
\theta_{3}({\cal T})^{\frac{c-1}{2}-4(\Delta_1+\Delta_2+\Delta_3+\Delta_4)}\times\\
\times
\Big(1+H_{non-classical}(\Delta_1,\Delta_2,\Delta_3,\Delta_4;\Delta|x)\Big),\ \ \ \ \
\ \ \ \
\quad {\cal T}=i\frac{K(1-x)}{K(x)}\ \ \ \ \
\ \ \ \
\label{Zaf}\nn
\ee

This is well-known to be the case
for the $SU(2)$ theory with
$N_f=4$ matter fundamentals with the masses such that $\Delta_i=\frac{1}{16}$,
$\Delta=\frac{a^2}{g^2}$
and the deformation parameters such that
$c=1$. Then:
\be
B\left(a|x\right)=\frac{e^{\frac{\pi i}{g^2} {\cal T} a^2}}{\left[x(1-x)\right]^{\frac{1}{8}}
\theta_3({\cal T})}
\\
\\
B\left(a|x\right)=\int\frac{db}{g}\, e^{\frac{2\pi i a b}{g^2}} B\left(b|1-x\right)
\ee

The second exactly solvable case is the $SU(2)$ theory with the adjoint matter field
with the adjoint mass and the deformation parameters such that $\Delta_{\mbox{\footnotesize{ext}}}=0$ and $c=1$.
This theory corresponds to
the toric conformal block (with $\Delta=\frac{a^2}{g^2}$) and
\be
B\left(a|{\cal T}\right)=\frac{e^{\frac{2\pi i}{g^2} {\cal T} a^2}}{\eta({\cal T})}\\
\\
B\left(a|{\cal T}\right)=\int \frac{db}{g}\, e^{\frac{4\pi i a b}{g^2}}B\left(b\left|-\frac{1}{{\cal T}}\right.\right)
\ee
in this case.

As one can see, in these both examples the asymptotic form of the modular matrix is exact. In fact,
these two cases are basically equivalent due to the correspondence between the one-point conformal
block on a torus and a conformal block on a sphere, see \cite{Pog} and s.6.1.

\subsection{Fourier transform from ${\cal U}_q(sl_2)$ algebra}

These exact formulas can be compared with those obtained for the Ponsot-Teschner kernel (\ref{toric_MK}).
For instance, when the external dimension in the toric conformal block goes to zero, the complicated
kernel (\ref{toric_MK}) is drastically simplified.
Naively one would expect from
eq.(\ref{toric_MK}) and using (\ref{per}) that, in this case,
\be\label{53}
M=2^{3/2}\int \frac{e^{4\pi i a'r}dr}{16\sinh\left(\frac{\pi (a+r)}{\epsilon_1}\right)\sinh
\left(\frac{\pi (a+r)}{\epsilon_2}\right)\sinh\left(\frac{\pi (a-r)}{\epsilon_1}\right)
\sinh\left(\frac{\pi (a-r)}{\epsilon_2}\right)}
\ee
In fact, the denominator in the integrand of (\ref{toric_MK})
has double poles, which merge in the limit of
$p_0\rightarrow i\frac{\epsilon}{2}$, and the integration contour is pinched between
these two poles. Hence, one has to deal with this limit more carefully:
\be
{\cal M}(a,a'|0)=\oint\lm_{r=-a} \frac{s(a+r+i\epsilon/2+i\lambda)}{s(a+r-i\epsilon/2-i\lambda)}
\frac{e^{4\pi i a'r}dr}{4\sinh\left(\frac{\pi (a-r)}{\epsilon_1}\right)\sinh
\left(\frac{\pi (a-r)}{\epsilon_2}\right)}+\nn\\
+\oint\lm_{r=a} \frac{s(a-r+i\epsilon/2+i\lambda)}{s(a-r-i\epsilon/2-i\lambda)}
\frac{e^{4\pi i a'r}dr}{4\sinh\left(\frac{\pi (a+r)}{\epsilon_1}\right)\sinh\left(\frac{\pi (a+r)}{\epsilon_2}\right)}
\ee
Since
\be
\frac{s(a+r+i\epsilon/2+i\lambda)}{s(a+r-i\epsilon/2-i\lambda)}=\prod\lm_{m,n\geq 0}\frac{(m+1/2)\epsilon_1+(n+1/2)\epsilon_2-i(a+r+i\epsilon/2+i\lambda)}{(m+1/2)\epsilon_1+(n+1/2)\epsilon_2+i(a+r+i\epsilon/2+i\lambda)}
\nn\times\\ \times\frac{(m+1/2)\epsilon_1+(n+1/2)\epsilon_2+i(a+r-i\epsilon/2-i\lambda)}{(m+1/2)\epsilon_1+(n+1/2)\epsilon_2-i(a+r-i\epsilon/2-i\lambda)}=\nn\\
=\prod\lm_{m,n\geq 0}\frac{(m+1)\epsilon_1+(n+1)\epsilon_2-i(a+r+i\lambda)}{m\epsilon_1+n\epsilon_2+i(a+r+i\lambda)}
\frac{(m+1)\epsilon_1+(n+1)\epsilon_2+i(a+r-i\lambda)}{m\epsilon_1+n\epsilon_2-i(a+r-i\lambda)}\sim\nn\\
\mathop{\sim}_{m,n=0}\ \ \frac{1}{(a+r)^2-\lambda^2}\ \ \mathop{\sim}_{\lambda\rightarrow 0}\ \ \delta(a+r)
\ee
one finally obtains
\be
{\cal M}(a,a'|0)\rightarrow \frac{\sqrt{2}\cos\left(4\pi i\frac{aa'}{\epsilon_1\epsilon_2}\right)}{\mu'(a)}
\ee
Thus, indeed, in this limit the modular transformation reduces just to the simple Fourier transform
as we already saw in s.\ref{Ex_cases}.

The Fourier transform can be also derived in the limit of
$\epsilon_{1,2}\rightarrow 0$. Assume that
$\im\Big(\frac{\epsilon_1}{\epsilon_2}\Big)>0$, then the asymptotics of the
quantum dilogarithm has different signs depending on the direction
\be\label{dilog_asym} s(z|\epsilon_1,\epsilon_2)\sim \left\{
\begin{array}{c}
                                         e^{\frac{\pi i}{2}z^2},\quad \mathop{\rm arg} \epsilon_1-\frac{\pi}{2}<\mathop{\rm arg} z< \mathop{\rm arg} \epsilon_2+\frac{\pi}{2} \\
                                         e^{-\frac{\pi i}{2}z^2}, \quad \mathop{\rm arg} \epsilon_2-\frac{3\pi}{2}<\mathop{\rm arg} z< \mathop{\rm arg} \epsilon_1-\frac{\pi}{2}
                                       \end{array}
 \right.
\ee
This leads to the limit for the modular kernel for $\epsilon_{1,2}\rightarrow 0$
\be
M(a,a_D)\sim \int dt \exp\left\{\frac{\pi i}{2\epsilon_1\epsilon_2}\left(2 t^2-(t-a_D)^2+(t+a_D)^2 -(t-a)^2+(t+a)^2\right)\right\}\sim e^{\frac{2\pi i}{\epsilon_1\epsilon_2}aa_D}
\ee
Unfortunately, this approach does not allow one to find the next corrections
to the modular matrix, since in the asymptotics
$\epsilon_{1,2}\rightarrow 0$ there is only quadratic term in the
exponential (\ref{dilog_asym}), so further
corrections look exponentially small and are not obtained within the asymptotic
perturbation theory \cite{Q-log}. At the same time, as we demonstrate in the next section,
the actual corrections are not like this (not exponentially small), and it is a question
(not answered in the present paper),
how (\ref{53}) should be effectively treated in order to reproduce them.

\section{Genus expansion in matrix models and corrections to the modular kernel}
\subsection{Loop equations}

In order to obtain a way to effectively generate manifest formulas for the $S$-duality,
we are going to apply the technique of matrix models which, in accordance with
the AGT conjecture, describe the conformal blocks and Nekrasov functions, see s.\ref{mamo}. This technique allows one to
calculate the modular kernel, i.e. to construct the $S$-duality transformations iteratively
using the matrix model loop equations.

Thus, we consider the multiple integral (\ref{mmint}). The standard way to
deal with matrix integrals is
to construct loop equations for the k-point resolvents \cite{loopeq,AMMEO}
\be
r_k(\xi_1,\ldots,\xi_k)=\left<\sum_i{1\over \xi_1-z_i}\sum_i{1\over \xi_2-z_i}\
\ldots\ \sum_i{1\over \xi_k-z_i}\right>
\ee
where the brackets denote the average w.r.t. to the measure (\ref{mmint})
\be
\int_{\gamma_i} dz_i \left(\prod\lm_{j>i} z_{ij}^{2\beta}\right)
\prod\lm_a (z_i-q_a)^{\frac{2b\alpha_a}{g}}
\ee
We also use the connected correlators and denote them
through $\rho_k(\xi_1,\ldots,\xi_k)$. For $k=1$ we additionally shift
\be
\rho_1(z)=\sqrt{\beta}\left(gr_1(z)+\sum_a{\alpha_a\over z-q_a}\right)
\label{60}
\ee
The loop equations for the
unconnected resolvents form the following set of equations
$$
\beta r_{k+1}(\zeta,\zeta,x_1,\ldots,x_{k+1})+(\beta-1) \pr_{\zeta} r_k(\zeta,x_1,\ldots,x_{k-1})+
\sum\lm_{a}\frac{2\sqrt{\beta}\alpha_a}{g}\frac{r_k(\zeta,x_1,\ldots,x_{k-1})-r_k(q_a,x_1,\ldots,x_{k-1})}{\zeta-q_a}+
$$
\vspace{-0.4cm}
\be
+\sum_j\pr_{x_j}\frac{r_{k-1}(x_1,\ldots,x_{k-1})-\left.r_{k-1}(x_1,\ldots,x_{k-1})\right|_{x_j=\zeta}}{x_j-\zeta}=0
\ee
One can reformulate it in terms of the connected resolvents which admits
"the genus expansion" in powers of
the coupling constant $g$. These equations are a bit more involved.
The first few of them are
\be
g^2\beta \rho_2(z)+g\left(\sqrt{\beta}-\frac{1}{\sqrt{\beta}}\right)\left(\pr \rho_1(z)
+\sqrt{\beta}\sum\lm_{a}\frac{\alpha_a}{(z-q_a)^2}\right)+\rho_1^2(z)-
\sum\lm_a\frac{\pr_{q_a}{\cal F}}{z-q_a} -\beta\left(\sum\lm_{a}\frac{\alpha_a}{z-q_a}\right)^2
=0\nn\\
g^2\beta \rho_3(z)+\frac{1}{2}g(\beta-1)\pr\rho_2(z)+2\rho_1(z)\rho_2(z)-
\sum\lm_a\frac{\pr_{q_a} \rho_1(z)}{z-q_a}+\frac{1}{2}\pr^2 \rho_1(z)=0\nn\\
2\rho_1(z)\rho_3(z)+2\rho_2^2(z)-\sum\lm_a\frac{\pr_{q_a}\rho_2(z)}{z-q_a}+(\hat L\rho_2)(z)=0\nn\\
2\pr^2 \rho_1(z)\rho_2(z)+2\pr \rho_1(z)\pr\rho_2(z)+2\rho_1(z)(\hat L\rho_2)(z)-2\sum\lm_a\frac{\pr_{q_a}\rho_1(z)}{(z-q_a)^3}+
\frac{1}{12}\pr^4 \rho_1(z)=0 \label{56}
\ee
where
\be
(\hat L\rho_2)(z)= \left.\pr_z\pr_w \rho_2(z,w)\right|_{w=z}
\ee
and ${\cal F}$ differs from $F$ by omitting the normalization factor $\prod_{a<b}(q_q-q_b)^{2\alpha_a\alpha_b/g}$
from (\ref{mmint}).

The matrix model ($\beta$-ensemble) partition function can be calculated
using the SW equations (\ref{mmSW}) for
its logarithm. The complex spectral curve is determined from the genus zero contribution to
the first equation of (\ref{56})
which involves only the 1-point resolvent $\rho_1^{(0)}(z)$ (\ref{60})
\be\label{sc}
y^2(z)=\rho_1^{(0)}(z)^2=\sum\lm_a\frac{\pr_{q_a}{\cal F}}{z-q_a} +\beta\left(\sum\lm_{a}\frac{\alpha_a}{z-q_a}\right)^2
\ee
while the differential is
 \be
\Omega_{g,\beta}=\rho_1(z)dz
\ee

\subsection{Modular kernel construction}

Similarly to the expansion of the prepotential into powers of the string coupling constant $g$, one
can perform this expansion for the modular matrix. Thus, we assume the following expansions
\be
F(a,x)=\sum \lm_{k=0}^{\infty} g^{2k}F_{k}(a,x)\\
M(a,b)=\exp\left(\frac{2\pi i a b}{g^2}+\sum\lm_{k=0}^{\infty}g^{2k}\mathfrak{m}_k(a,b)\right)
\ee
which should be inserted into the definition of the modular kernel $M(a,b)$
\be
e^{\frac{2\pi i}{g^2}F(b,1-x)}=\int \frac{da}{g} M(a,b) e^{\frac{2\pi i}{g^2}F(a,x)}
\ee
Then, a simple computation leads to the following relation
\be
F(b,1-x)=\left(F(a_0,x)+a_0 b\right)+g^2\left(\mathfrak{m}_0(a_0,b)+F_1(a_0|x)-\frac{1}{4\pi i}\log F_0''(a_0|x)\right)+\nn\\+g^4\left(\mathfrak{m}_1(a_0,b)+F_2(a_0|x)+\frac{i}{4\pi}\frac{F_1''(a_0|x)}{F_0''(a_0|x)}+\frac{i}{4\pi}\frac{\mathfrak{m}_0''(a_0,b)}{F_0''(a_0|x)}-\frac{1}{2}\frac{(\mathfrak{m}_0'(a_0,b))^2}{F_0''(a_0|x)}-\frac{1}{2}\frac{(F_1'(a_0|x))^2}{F_0''(a_0|x)}+\right.\nn\\
\left. -\frac{\mathfrak{m}_0'(a_0,b)F_1'(a_0|x)}{F_0''(a_0|x)}-\frac{1}{32\pi^2}\frac{F_0^{(IV)}(a_0|x)}{(F_0''(a_0|x))^2}-\frac{i}{4\pi}\frac{F_0'''(a_0|x)F_1'(a_0|x)}{(F_0''(a_0|x))^2}+\frac{5}{96\pi^2}\frac{(F_0'''(a_0|x))^2}{(F_0''(a_0|x))^3}\right)+\ldots,
\ee
where $a_0$ is determined from the equation $F_0'(a_0|x)+b=0$ and the prime means the derivative w.r.t. $a_0$.
This relation can be also viewed as defining the function $x(a,b)$ in an implicit way
\be
F_0'(a|x(a,b))=-b
\ee
Using this formula, one can eliminate the $x$-dependence from the modular kernel.
For instance, in the first order one has
\be
\mathfrak{m}_0(a,b)=F_1(b,1-x(a,b))-F_1(a|x(a,b))+\frac{1}{4\pi i}\log F_0''(a|x(a,b))
\ee

\subsection{First few terms of expansion}

Here we present some explicit expressions for the first few terms of expansion of the
prepotential calculated on the lines of s.5.1. Then, following s.5.2, we find
the corresponding genus expansion of the modular kernel.

\paragraph{Example 1: \underline{$\epsilon_1=-\epsilon_2=g$ or
$\beta=1$ with $\mu_1=2t m_1,\mu_{2,4}=0,\mu_3=2 tm_3$.}}

In this case, the prepotential has the following expansion
\be
F(\mu_1=2t m_1,\mu_2=0,\mu_3=2 tm_3,\mu_4=0|x)=\sum\lm_{k,m=0}^{\infty} I_{k,2m}(x) g^{2k} t^{2m}
\ee
Several first terms in this expansion are given by the following expressions
\be
I_{00}=-\frac{\pi a^2 K(1-x)}{K(x)}\\
I_{02}=2(m_1^2+m_3^2)\log a\\
I_{04}=-\frac{2(m_1^4+m_3^4)}{3\pi^2 a^2}\left((x-2)K(x)+3E(x)\right)-\frac{4m_1^2m_3^2}{\pi^2a^2}K(x)\left((x-1)K(x)+E(x)\right)\\
I_{10}=-\frac{1}{2}\log a\\
I_{12}=\frac{2(m_1^2+m_3^2)}{3\pi^2 a^2}K(x)\left((x-2)K(x)+E(x)\right)\\
I_{14}=\frac{4(m_1^4+m_3^4)}{3\pi^2 a^4}K^2(x)\left[(x^2-3x+3)K^2(x)+4(x-2)K(x)E(x)+6E^2(x)\right]+\nn\\
+\frac{8m_1^2 m_3^2}{3\pi^4 a^4}\left[(3x^2-7x+4)K^2(x)+2(4x-5)K(x)E(x)+6E^2(x)\right]\\
I_{20}=-\frac{K(x)}{8\pi^2 a^2}\left((x-2)K(x)-3E(x)\right)\\
I_{22}=-\frac{m_1^2+m_3^2}{60\pi^4 a^4}K^2(x)\left((48 x^2-143x+143)K^2(x)+190(x-2)K(x)E(x)+285 E^2(x)\right)
\ee
\be
I_{24}=-\frac{K^3(x)(m_1^4+m_3^4)}{90\pi^6 a^6}\left(-1646 K^3(x)+4350 E^3(x)+2469 x K^3(x)-8700 K(x) E^2(x)+6476 K^2(x) E(x)\nn-\right. \\ \left. -1783 K^3(x)^3 x^2+480 x^3 K^3(x)+4350 K(x) E^2(x) x+2126 K^2(x) x^2 E(x)-6476 K^2(x) x E(x)\right)-\nn\\
-\frac{K^3(x)m_1^2m_3^2}{45\pi^6 a^6}\left(-1646 K^3(x)+4350 E^3(x)+2469 K^3(x) x-8700 K(x) E^2(x)+6476 K^2(x) E(x)-1783 K^3(x) x^2+\right.\nn\\ \left.+480 x^3 K^3(x)+4350 K(x) E^2(x) x+2126 K^2(x) x^2 E(x)-6476 K^2(x) E(x) x\right)
\ee
Surprisingly, in this case of only two zero masses
the modular matrix does not seem to differ from the Fourier transform:
\be
\boxed{e^{\frac{1}{g^2}F(x|a)}=\int \frac{d b}{g} e^{\frac{2\pi i ab}{g^2}+{\cal O}(t^6,g^4)}e^{\frac{1}{g^2}F(1-x|b)}}
\ee
Even more surprisingly, this seems to remain true for arbitrary masses: corrections are absent
whenever $\beta=1$ (though we do not possess yet a complete evidence in favor of this conjecture).

\paragraph{Example 2: \underline{Double deformation $\beta\ne 1$
with $\mu_1=\mu_2=-\mu_3=-\mu_4=\epsilon/2$.}}

In the case of the arbitrary double deformation
the modular kernel is no longer the Fourier transform. We
demonstrate this in the simplest case of  masses $\mu_1=\mu_2=-\mu_3=-\mu_4=\epsilon/2$.
Then, the prepotential is
\be
F=-\frac{\pi a^2 K(1-x)}{K(x)}+\frac{g^2}{2}\left(3\beta-7+\frac{3}{\beta}\right)\log a-\nn\\
-\frac{g^4}{8a^2}\frac{\left(3\beta-7+\frac{3}{\beta}\right)\left(\beta-3+\frac{1}{\beta}\right)}{\pi^2}K(x)
\left((x-2)K(x)+3E(x)\right)+\ldots
\ee
The modular matrix in this case reads
\be
\boxed{M(a,b)=\exp\left(\frac{2\pi i a b}{g^2}+\frac{3(\beta-1)^2}{2\beta}\log\frac{a}{b}-\frac{3i g^2}{16\pi \beta^2}\frac{(\beta-3)(3\beta-1)(\beta-1)^2}{a b}+\ldots\right)}
\ee
Thus, for $\beta\ne 1$ corrections to the Fourier transform are non-vanishing.

\bigskip

\section{$S$-duality in the NS limit: comments and remarks}

In the previous section we used the matrix model technique in order
to manifestly construct the $S$-duality kernel as a series in powers
of $\beta -1$, i.e. around the line $\beta=1$. In this section we
briefly discuss peculiarities of another important line $\epsilon_2=0$ (the NS limit
\cite{NS}) which are essential for further studies, since
in the NS limit the modular properties are much simpler than in the generic case,
and there is a hope to understand them in much more details. This limit is also interesting
from the point of view of integrable systems, because there the remaining
$\epsilon_1$ can be treated as
a quantization parameter.

\subsection{Modular transformation on torus\label{Pog}}

So far using the matrix model technique we discussed how to obtain the $S$-duality kernel
of conformal blocks on a sphere. One can similarly deal with the conformal blocks
on a torus, though the matrix model is somewhat more involved \cite{MMtorus}.
In practice, it is simpler to use an equivalence of the one-point
toric conformal block and a special four-point spherical conformal block \cite{Pog}.
However, in \cite{Pog} this equivalence is established
through a sophisticated recursion procedure which is still not very well understood.
Here we describe this equivalence in a much more explicit way, but only
in the limit $\epsilon_2\rightarrow 0$, i.e. when there is an underlying quantum
integrable system (the elliptic Calogero model
\cite{SWint,BS,BSmore}) and a "quantized" SW curve. In order to completely describe
the conformal block in this limit, one has to insert an additional field, degenerate
at the second level \cite{SO}, and consider the equation for the conformal block that emerge
after this \cite{BPZ}. This Schr\"odinger-Baxter equation is exactly what one calls
the "quantized" SW curve, and the
logarithmic derivative of solution of this equation is the SW differential. Using this
SW data, one constructs the Nekrasov function in the NS limit, $\epsilon_2\to 0$.

Thus, the
conformal blocks to compare are the 5-point spherical block and
the toric 2-point block, both with one of the fields degenerate at
the second level and the properly adjusted intermediate dimensions
\cite{surop}. We demonstrate that the differential equation for the toric
block reduces to a particular case of the spherical block
differential equation. The original four-point and one-point conformal blocks can
afterwards be extracted from the asymptotics of the solutions
by a procedure presented in \cite{surop}.

In the limit of $\epsilon_2\to 0$, the differential equation for the
toric conformal block is \cite{surop}
\be
\left\{2\pi
i\frac{\pr}{\pr\tau_0}-4\frac{\epsilon_1}{\epsilon_2}
\left[\sqrt{(X-e_1)(X-e_2)(X-e_3)}\pr_X\right]^2+
\frac{\mu(\epsilon_1-\mu)}{\epsilon_1\epsilon_2}X\right\}\Psi\Big(\tau_0,X\Big)=0
\ee
where, as compared with \cite[eq.(36)]{surop}, we left only the
terms essential in the limit $\epsilon_2\rightarrow 0$ and
introduced the variable $X=\wp(z|\tau_0)$, i.e.
$z=\frac{1}{2}\int\lm_{X}^{\infty}\frac{ds}{\sqrt{(s-e_1)(s-e_2)(s-e_3)}}$,
with
\be
e_1=\frac{\pi^2}{3}(2-x)\theta_3^4(\tau_0), \quad e_2=-\frac{\pi^2}{3}(1+x)\theta_3^4(\tau_0),
\quad e_3=-\frac{\pi^2}{3}(1-2x)\theta_3^4(\tau_0),
\quad x=\frac{\theta_2^4(\tau_0)}{\theta_3^4(\tau_0)}
\ee
The external dimension
$\Delta=\frac{\mu(\epsilon_1-\mu)}{\epsilon_1\epsilon_2}$
parameterizes the mass $\mu$ of the adjoint matter hypermultiplet
(adding this hypermultiplet corresponds to the toric conformal block
in the AGT framework).

Now, change of the variable $X=e_2-t(e_1-e_2)$ leads to the equation
\be
\left\{\frac{1}{2}x(x-1)\pr_x+t(t-1)(t-x)\left[\pr_t^2-\frac{1}{2}
\left(\frac{1}{t}+\frac{1}{t-1}+\frac{1}{t-x}\right)\right]+
\frac{\mu(\epsilon_1-\mu)}{4\epsilon_1\epsilon_2}
\left(t-\frac{1}{3}(1+x)\right)\right\}\Psi(x,t)=0
\ee
By rescaling the wave function
$\tilde{\Psi}(t)=\left[t(t-1)(t-x)\right]^{-\frac{1}{4}}\Psi(t)$,
one can eliminate the term linear in $\pr_t$, up to the order
$\mathcal{O}(\epsilon_2^0)$. This
procedure redefines the SW integral, with the intermediate dimension
$\Delta_\alpha=\frac{\alpha(\epsilon-\alpha)}{\epsilon_1\epsilon_2}$),
in accordance with
$$
\frac{\epsilon_1}{2\pi i}\oint_A d_t\log\tilde\Psi=\frac{\epsilon_1}{2\pi i}\oint_A d_t\log\Psi-
\frac{\epsilon_1}{2}=\alpha-\frac{\epsilon_1}{2}=a
$$
This equation
coincides with that for the spherical conformal block
\cite[eqs.(85)-(86)]{surop} in the limit $\epsilon_2\rightarrow 0$
provided one rescales the modular parameter $\tau_0\rightarrow
2\tau_0$ and the wave function
$\psi=\tilde\Psi\left[x(x-1)\right]^{-\frac{\Delta}{12}}$:
\be
\left[x(x-1)\pr_x-\frac{\epsilon_1}{\epsilon_2}t(t-1)(t-x)\pr_t^2-
\frac{\mu(\epsilon_1-\mu)}{4\epsilon_1\epsilon_2}(t-x)\right]\psi (x,t)=0
\ee
Since the wave functions are the same for toric and spherical cases, the same is true for the
conformal blocks.
Thus, we reproduce the answer of \cite{Pog}, but only in the limit
$\epsilon_2\rightarrow 0$
\be Z_{{\rm
sphere}}\left(\Delta_{1,2,3}=0,\Delta_4=\frac{\Delta}{4}\Big|x=
\frac{\theta_2^4(2\tau_0)}{\theta_3^4(2\tau_0)}\right)=
\eta^{-2\Delta}(\tau_0)\left(\frac{\theta_3^8(2\tau_0)}{\theta_2^4(2\tau_0)
\theta_4^4(2\tau_0)}\right)^{\frac{\Delta}{12}}Z_{{\rm
torus}}(\Delta|\tau_0)
\ee
This means that the toric conformal block
has the same transformation properties as the spherical one, its
modular/$S$-duality operator in the limit $\epsilon_{1,2}\to 0$
being
\be
M(a,a')=e^{\frac{4\pi i aa'}{\epsilon_1\epsilon_2}}
\ee
Note that we made the rescaling $\tau_0\rightarrow2\tau_0$,
thus this formula is in accordance with the results of \cite{PT}. In sect.4.2 we already
encountered an explicit application of this result.

\subsection{$S$-transformation of effective coupling constants\label{bvse}}

Throughout the paper we mostly concentrated on the modular kernel for the transformation
$x\to 1-x$, which turns to resemble the Fourier transform.
One may also ask how the effective coupling constant behaves under this transformation.

As we discussed above, the $S$-duality transformation of an
effective charge matrix defined as
$||{\cal T}_{\epsilon_1,\epsilon_2}||_{i,j}=\frac{\pr^2
F_{\epsilon_1,\epsilon_2}}{\pr a_i\pr a_j}$, has the simple form in
the SW case:
\be
\hat{S}({\cal T}_{0,0})=-({\cal T}_{0,0})^{-1}
\ee
In fact, it is
preserved in the once deformed case too:
\be
\hat{S}({\cal T}_{\epsilon_1,0})=-({\cal T}_{\epsilon_1,0})^{-1}
\ee
The reason
that this property survives the deformation is that in this case
there still exists the closed (Baxter or Schr\"odinger) equation for
the SW differential so that one can immediately construct the
Bohr-Sommerfeld integrals to obtain the SW prepotential, and the SW differential
remains intact under duality transformations. However,
this is no longer the case when the both deformations are switched
on. Indeed, in this case the SW differential is determined only from
the loop equations involving all multi-point resolvents and
\be
 \hat{S}({\cal T}_{\epsilon_1,\epsilon_2})\neq -({\cal T}_{\epsilon_1,\epsilon_2})^{-1}
\ee
There is the following technical reason for this. Consider the
space of coupling constants $x=e^{2\pi i\tau_0}$ and the fiber bundle
of the Riemann surfaces over this space $u(x)=\langle\Tr
\phi^2\rangle$. With this bundle, one can construct the A- and
B-periods of the 1-resolvent: $a(x)=\langle\phi\rangle$, $b(x)$. As
soon as the SW differential or the equation for it in the once
deformed case depends only on $u(x)$, but not on its derivatives,
one can locally interchange the cycles at some point $x_0$, i.e.
just change the value of $a(x)$ at this point $x=x_0$, and it is
completely independent on the value of $a(x)$ at any other point
$x$. This property is broken, however, by the matrix model
corrections, since then the derivatives of $u(x)$ manifestly enter
the differential (and the equations for it). On the other hand, we
fix $a(x)=const$, $b(x)$ is a non-trivial function and one can no
longer permute the cycles locally at some $x_0$. To put it
differently, in order to completely fix the multi-resolvents in
matrix model, one has to impose some condition on their A-periods: for instance, to
require that they vanish. This
manifestly breaks the symmetry under permutations of the A- and
B-cycles, see Fig.2.
\begin{figure}[h]
\begin{center}
\includegraphics[scale=0.5]{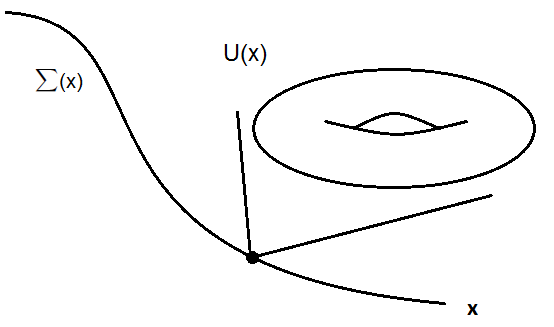}
\end{center}
\label{torus_fam} \caption{Family of effective tori defining the
matrix model partition function}
\end{figure}

\subsection{On the full duality group\label{cdg}}

We discussed so far only the $S$-transformation from the duality group. One may ask
what is the full duality group, i.e. when one adds the generator of the
$T$-transformation. This question was addressed in detail in numerous papers
\cite{SW1,SW2,Sdua2,Arg}, still a few short remarks deserve mentioning.

Note that so far we considered different groups of transformations. First of all, it is
the group of modular transformations of the conformal blocks. It is related by the AGT
relations with the $S$-duality group. The modular transformations are
induced by the third group, which relates different possibilities to arrange the brackets in
the tensor products. This group sometimes coincides with the group of permutations of points,
as we discuss in this section below.

\subsubsection{$SL(2,\mathbb{Z})$ and Racah identities}

We start with the case of $SU(2)$ gauge group, i.e. with the four-point conformal block.
Note that the associativity of the operator algebra implies an essential property of
the Racah-Wiegner coefficients:
considering multiple products of representations and
rearranging brackets in different ways, one arrives at commutative diagrams.
One of these is
\be\label{pentagon}
\begin{CD}
(T_1\otimes T_2)\otimes T_3 @>{\hat{T}}>> (T_2\otimes T_1)\otimes T_3 @>{\hat{R}}>> T_2\otimes (T_1\otimes T_3)\\
@VV{\hat{R}}V   @. @VV{\hat{F}}V\\
T_1\otimes(T_2\otimes T_3) @<{\hat{T}}<< T_1\otimes (T_3\otimes T_2) @<{\hat{R}}<< (T_1\otimes T_3)\otimes T_2
\end{CD}
\ee
which imposes a nontrivial restriction $\hat{T}\hat{R}\hat{F}\hat{R}\hat{T}=\hat{R}$
on the modular kernel.
For the modular and $S$-duality transformation with generators $\hat S$ and $\hat T$ one has
$\hat{R}=\hat{T}\hat{S}$, $\hat{F}=\hat{1}$, and the Racah relation is equivalent to
$(\hat{T}\hat{S})^3=1$. Along with $\hat{S}^2=1$ this implies that these transformations
generate $SL(2,\mathbb{Z})$.

As we discussed in the previous subsection,
one can realize these generators in the NS limit
either in terms of the effective coupling constant ${\cal T}$
on the gauge theory side ($S$-duality) as
\be
\hat{S}: \quad {\cal T}\mapsto -1/{\cal T}, \quad \hat{T}: \quad {\cal T}\mapsto {\cal T}+1,
\ee
or in terms of the double ratio (bare coupling constant in the gauge terms)
of four points on the CFT side (modular transformation) as
\be
\hat{\bar S}: \quad x\mapsto 1-x,\quad \hat{\bar T}: \quad x\mapsto \frac{x}{x-1}
\ee
In the first representation, one obtains only two relations for the generators:
$\hat S^2=1$ and $(\hat{T}\hat{S})^3=1$, while in the second one there is the additional
restriction $\hat{\bar T}^2=1$, i.e. the generators, in this case, form a finite group of
permutations $S_3$. This seeming contradiction with the AGT relation is resolved once one
notice that the main object of our interest, the conformal block also does not
obey the relation $\hat{T}^2=1$ because of the singular behaviour:
$B\sim x^{\Delta-\Delta_1-\Delta_2}$, $x\rightarrow 0$. As a result, the action of $\hat{T}^2$
which carries the point $q$ around zero gives rise to a non-trivial
monodromy factor, thus, the conformal block provides the representation of $SL(2,\mathbb{Z})$,
not of $S_3$.

\subsubsection{Multi-point conformal blocks}

This difference between the permutation and duality groups becomes more profound for
multi-point conformal blocks. Indeed, in this case not all possible modular transformations
(i.e. all possible ways to place brackets in the tensor products) can be associated with
permutations of points. This happens starting from the six-point conformal
block\footnote{It is interesting to note that similarly an absolutely new, more complicated
behaviour characterizes the $n$-point gluon amplitudes within the Alday-Maldacena framework
at $n\ge 6$ as compared with $n=4,5$ \cite{MMT}.
}.
In the case of
the five-point conformal block this is still possible, and the modular transformations
can be described by the $S_4$ group of permutations. The duality group in this case also
has nothing to do with the permutation group. The five-point conformal block
described theory with the gauge group $SU(2)\times SU(2)$,
however, the duality group is not a direct product of two $SL(2,\mathbb{Z})$, see the detailed
discussion in \cite{Arg}. It is done there for SW theory, i.e. for $\epsilon_1,\epsilon_2=0$,
however, it is enough
in this case, though one can definitely repeat the matrix model
calculations for arbitrary $\epsilon_1$ and $\epsilon_2$.\footnote{A counterpart
of the spectral curve (\ref{sc}) in the case of generic multi-point conformal block is
in the massless limit
$$
y^2(z)=\sum\lm_i \frac{x_i(x_i-1)\pr_{x_i}{\cal F}}{z(z-x_i)(z-1)}
$$
and the SW (genus zero) differential is
$$
\Omega=y(z)dz
$$
The bare coupling constants are related with the double ratios in this case via formulas like
$x_k=\exp\left(2\pi i \sum\lm_{i=1}^{N-k}\tau_0^i\right)$.}

The S-duality group contains two $SL(2,\mathbb{Z})$ subgroups and an additional generator
mixing them, i.e. one can totally extract five generators: $\hat{S}_{1,2}$, $\hat{T}_{1,2}$
and the mixing generator $\hat{Q}$. As previously,
one can construct the Racah relation, which in this case reads
\be
\hat{R}_1\hat{T}_1 \hat{R}_2\hat{R}_1=\hat{Q}\hat{R}_1\hat{R}_2
\ee
with the Racah coefficients given by $\hat{R}_{1,2}=\hat{T}_1\hat{T}_2\hat{S}_{1,2}$.
Since the $S$-duality group contains $SL(2,\mathbb{Z})$ subgroups,
the relation $\left(\hat{S}\hat{T}\right)^3=1$ still holds.

As we mentioned, beginning with the six-point case
the modular transformations are not reduced to permutations of points. The reason is that
the new diagram emerges, and one has to describe within the modular transformation
the map
\be
\begin{picture}(50,14)\put(0,0){\line(1,0){50}}
  \put(10,0){\line(0,1){10}}
  \put(20,0){\line(0,1){10}}
  \put(30,0){\line(0,1){10}}
  \put(40,0){\line(0,1){10}}
  \put(1,2){\makebox(0,0)[cc]{$0,\Delta_1$}}
  \put(7,12){\makebox(0,0)[cc]{$x_1,\Delta_2$}}
  \put(17,12){\makebox(0,0)[cc]{$x_2,\Delta_3$}}
  \put(27,12){\makebox(0,0)[cc]{$x_3,\Delta_4$}}
  \put(37,12){\makebox(0,0)[cc]{$1,\Delta_5$}}
  \put(50,2){\makebox(0,0)[cc]{$\infty,\Delta_6$}}
  \end{picture}
\qquad \longrightarrow\qquad
\begin{picture}(50,14)\put(0,0){\line(1,0){50}}
  \put(10,0){\line(0,1){10}}
  \put(25,0){\line(0,1){5}}
  \put(25,5){\line(1,1){7}}
  \put(25,5){\line(-1,1){7}}
  \put(40,0){\line(0,1){10}}
  \put(1,2){\makebox(0,0)[cc]{$0,\Delta_1$}}
  \put(7,12){\makebox(0,0)[cc]{$x_1,\Delta_2$}}
  \put(17,14){\makebox(0,0)[cc]{$x_2,\Delta_3$}}
  \put(33,14){\makebox(0,0)[cc]{$x_3,\Delta_4$}}
  \put(43,12){\makebox(0,0)[cc]{$1,\Delta_5$}}
  \put(50,2){\makebox(0,0)[cc]{$\infty,\Delta_6$}}
  \end{picture}\nn
\ee
This implies that the modular and $S$-duality groups are different for these theories:
this transformation can not be described within the framework of the same
SYM theory with modified parameters. This issue requires further efforts to be understood.

\section{Conclusion}

In this paper, we reviewed the problem of constructing the fusion coefficients in $2d$
conformal field theory from the point of view of
the AGT conjecture. We provided some explicit formulas relating the modular map of the conformal blocks
to the $S$-duality map of the
$\Omega$-background-deformed prepotential or of the Nekrasov partition functions.
We demonstrated that $S$-duality transformation is actually the Fourier transform at $\beta=1$,
but acquires non-trivial corrections for $\beta\ne 1$. These corrections can be found
order-by-order from the Virasoro constraints (loop equations) for the "conformal"
$\beta$-ensemble (i.e. with logarithmic potential) of \cite{AGTmamo,MMCB}.

Still more effort is needed to reproduce this result from the
${\cal U}_q(sl_2)$-algebra consideration by B.Ponsot and J.Teschner.
A generalization to multi-point conformal blocks, deformed Virasoro
algebras and ${\cal W}_N$-algebras are also open problems.

\section*{Acknowledgements}

Our work is partly supported by Ministry of Education and Science of
the Russian Federation under contract 14.740.11.0081, by NSh-3349.2012.2,
by RFBR grants 10-02-00509 (D.G. and A.Mir.), 10-02-00499 (A.Mor.) and
by joint grants 11-02-90453-Ukr, 12-02-91000-ANF, 12-02-92108-Yaf-a,
11-01-92612-Royal Society.

\end{document}